\documentclass[runningheads]{llncs}
\usepackage{graphicx}
\usepackage[misc]{ifsym}
\begin{document}
\title{Liver Segmentation from Multimodal Images using HED-Mask R-CNN }
\author{Supriti Mulay\inst{1,2}\textsuperscript{(\Letter)}\orcidID{0000-0002-2115-3642} \and Deepika G\inst{3} \and Jeevakala S\inst{2} \and Keerthi Ram\inst{2} \and Mohanasankar Sivaprakasam\inst{1,2}}

\institute{Indian Institute of Technology Madras, Chennai, India
\newline \email{supriti@htic.iitm.ac.in} \and
Healthcare Technology Innovation Centre, Chennai, India \and
Manipal Institute of Technology, Manipal, India\\
}

\maketitle              
\begin{abstract}
Precise segmentation of the liver is critical for computer-aided diagnosis such as pre-evaluation of the liver for living donor-based transplantation surgery. This task is challenging due to the weak boundaries of organs, countless anatomical variations, and the complexity of the background. Computed tomography (CT) scanning and magnetic resonance imaging (MRI) images have different parameters and settings. Thus, images acquired from different modalities differ from one another making liver segmentation challenging task. We propose an efficient liver segmentation with the combination of holistically-nested edge detection (HED) and Mask- region-convolutional neural network (R-CNN) to address these challenges. The proposed HED-Mask R-CNN approach is based on effective identification of edge map from multimodal images. The proposed system firstly applies a preprocessing step of image enhancement to get the 'primal sketches' of the abdomen. Then the HED network is applied to enhanced CT and MRI modality images to get better edge map. Finally, the Mask R-CNN is used to segment the liver from edge map images. We used a dataset of 20 CT patients and 9 MR patient from the CHAOS challenge. The system is trained on CT and MRI images separately and then converted to 2D slices. We significantly improved the segmentation accuracy of CT and MRI images on a database with Dice value of 0.94 for CT, 0.89 for T2-weighted MRI and 0.91 for T1-weighted MRI.

\end{abstract}
\section{Introduction}
The liver is the largest digestive gland and detoxification organ in the human body. A CT and MRI are used to detect any injury or bleeding in the abdomen. This is a painless and accurate way to detect an internal trauma which helps in saving patients’ lives. 
Automatic medical image segmentation approaches that are introduced in the last two decades have been the most successful methods for medical image analysis. The feasibility of a CNN to be generalized to perform liver segmentation across various imaging strategies and modalities is used in \cite{Wang}. Patrick et al \cite{Christ} presented a method to automatically segment liver and lesions in CT and MRI abdomen images using cascaded fully convolutional neural networks (CFCNs) enabling the segmentation of large-scale medical trials and quantitative image analysis. Zhe Liu et al \cite{LiuZhe} proposed liver sequence CT image segmentation solution GIU-Net, which consolidates an improved U-Net and a graph cutting algorithm, to take care of the low contrast between a liver and its surrounding organs issue. The problem of the large difference among individual livers in CT image was also addressed in \cite{LiuZhe}. 

The principal approach to image segmentation is to detect image discontinuities, edges are one of those. Canny edge detection is the most popular technique for edge detection but has limitations that different scales not directly connected, also exhibit spatial shift and inconsistency \cite{Xie}. 

HED was proposed by Saining Xie et al. \cite{Xie} to address these limitations. The original HED network was intended for edge discovery purposes in normal pictures, which catches fine and coarse geometrical structures (e.g. contours, spots, lines, and edges), while we are keen in capturing 'primal structure' in abdomen images. 

We chose holistically-nested edge detection because it addresses the challenging ambiguity in edge and object boundary detection significantly. We proposed a unique method that can perform segmentation of liver on various modalities in detecting features and instance segmentation with a holistically nested edge (HED)-Mask R-CNN. We investigate a deep learning methodology that jointly detect the edges and then segments the liver. The network is trained on a subset of the CHAOS challenge and evaluated on other subset data of CHAOS challenge for both CT and MRI modalities. 

Our contributions in the present work are,
\begin{itemize}
    \item use of enhancement method to get 'primal sketches' of abdomen images 
    \item utilize holistically-nested edge Mask RCNN (HED-Mask R-CNN) to get edge map
    \item applying the Mask R-CNN to segment liver from edge map images.
    \item lastly, we demonstrate the generalization and adaptability of HED-Mask R-CNN to different modalities 
\end{itemize}

The remainder of the paper is described in following subsections. The Section $2$ deals with joint network approach , Section $3$ include the experiment and results, Section $4$  presents a discussion of the proposed method and finally, Section $5$ draws the conclusions of this work. 

\section{Joint Network Approach}
The segmentation process of liver consists of joint deep learning pipeline:   pre-processing, edge map detection (Fully convolutional network (FCN) with deep supervision) \cite{Xie}, feature extractor with fine-tuning layers \cite{He}, as depicted in Fig. 1.

\begin{figure}[ht]
    \centering
    \begin{center}
    \begin{tabular}{c} 
    \includegraphics[scale = 0.5]{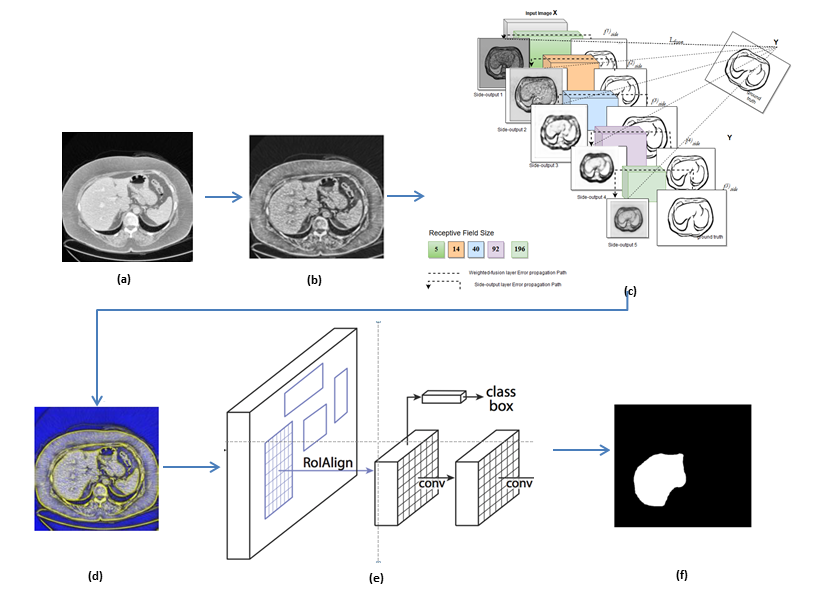}
    \end{tabular}
    \end{center}
    \setlength{\abovecaptionskip}{0.0pt}
    \setlength{\belowcaptionskip}{0.0pt}
    \caption{Joint network architecture for automated semantic liver segmentation.
    a) Original image b) enhanced image c) HED network  d) The original image multiplied by the obtained edge map e) Mask R-CNN. f) Segmented liver output.}
    {\label{fig:example1}
}
\end{figure}  

\subsection{Pre-processing}
Each DICOM slice is converted to PNG image and then pre-processing is carried out. Image noise, spatial resolution, and slice thickness affect CT and MRI images. An image enhancement technique is firstly applied, to get the organ sketches in abdomen images. We applied a separate enhancement technique to CT and MRI images because of their different resolutions. The CT images have been enhanced by modified sigmoid adaptive histogram equalization algorithm \cite {Supriti}. An adaptive histogram equalization (CLAHE) and sigmoid function are applied to preserve the mean brightness of the input CT images. We apply unsharp contrast enhancement filter to allow better differentiation of abnormal liver tissue in the case of MRI images. The abdomen organ features are enhanced prominently by this method. Fig.2(a) and 2(c) shows original images of CT and MRI respectively and Fig. 2(b) and 2(d) shows the enhanced images.

\begin{figure}[ht]
    \centering
    \begin{center}
    \begin{tabular}{c} 
    \includegraphics[height= 3cm]{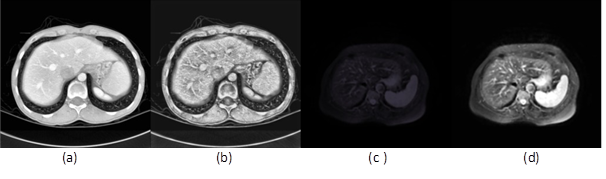}
    \vspace{-2mm}
    \end{tabular}
    \end{center}
    \caption{(a) and (c) CT and MRI  original images, (b) and (d) corresponding enhanced images}
    {\label{fig:example2}
    
    \vspace*{-4mm}
}
\end{figure}

\subsection{Holistically-nested edge detection approach}
Deep supervision used in HED that accounts for low-level predictions resulting in better edge map, is one of the reasons to consider HED in our method. We thus chose HED that automatically learns rich and important hierarchical representations from MRI/CT images to resolve the challenging ambiguity in edge and object boundary detection. It incorporates multi-scale and multi-level learning of deep image features utilizing auxiliary cost functions at each convolutional layer. This network architecture is with 5 stages, including strides of 1, 2, 4, 8 and 16, can capture the inherent scales of organ contours \cite{Roth}. Consequently, HED-based profound system models have been effectively utilized in medical image analysis for brain tumor segmentation \cite{Ying}, prostate segmentation \cite{Cheng}, pancreas localization, and segmentation \cite{Roth}, retinal blood vessel segmentation \cite{Y.Lin}.

The  network  structure  is  initialized based  on  an  ImageNet  pre-trained  VGGNet  model. The enhanced images are fed as an input to HED to get refined edge map. Organ edge/interior map predictions can be obtained at each side-output layer. The refined edge maps produced as side output are considered as an input to our next network.  Superior output for each modality is chosen. A side output $6$ is chosen for CT images whereas side output $0$ for MRI images.  Fig.3  shows the CT/MRI edge maps chosen to train the next CNN network for segmentation.

\begin{figure}[ht]
    \centering
    \begin{center}
    \begin{tabular}{c} 
    \includegraphics[scale = 0.5]{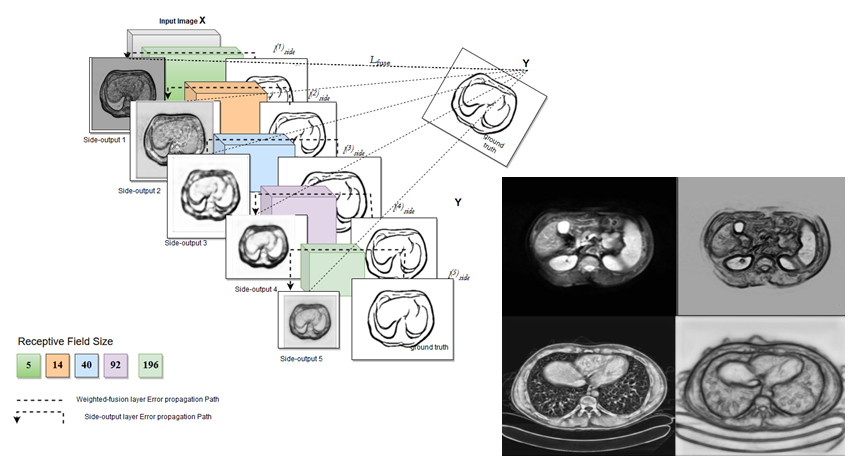}
    \end{tabular}
    \end{center}
    \caption{CT and MR images with associated edge map images of training data}
    {\label{fig:example3}
}
\end{figure} 

\subsection{Segmentation with Mask R-CNN}
\setcounter{secnumdepth}{3}
\subsubsection{Data preparation}

We use popular geometric augmentation techniques, flipping, and sharpening in this work. Elastic deformation distorts images locally by moving individual pixels around following a distortions field with strength sigma. We applied elastic distortion with alpha in range 0.5 to 3.5 and a smoothness parameter of 0.4.

\subsubsection{Mask R-CNN}
One of the most successful deep learning network for image segmentation is Mask R-CNN\cite{He}. Therefore we chose this model to segment the liver from both modalities. Enhanced image multiplied with obtained edge map from HED is used as input for Mask R-CNN to understand the 3D structure. Prior knowledge of edge map gives advantage in segmenting objects with a large variety in appearance and lack of texture to strong textures. We have used an end-to-end pre-trained Mask R-CNN model with a Resnet-101-FPN backbone in this study. This model has been pre-trained on Imagenet dataset.  It predicts the masks of detected regions and classifies them into one of the classes given at the time of training. We choose an existing open-source implementation \cite{matterport} using Tensorflow deep learning framework.

The inputs for HED FCN are gray-scale images of size $512\times 512 $ and their outputs are images of size $512 \times 512 \times 3$. The model is implemented in Keras\footnote[1]{https://keras.io/} with the TensorFlow \footnote[2]{https://www.tensorflow.org/} backend.

\subsubsection{Training Strategy}
Even though Mask R-CNN is profound enough and is equipped for learning appropriate parameters for liver segmentation, it is inclined to over-fitting issues. We utilize an effective technique such as Adam optimizer \cite{Kingma} to alleviate this issue and boosting the training. 

The multi-task loss function of Mask R-CNN combines the loss of classification, localization and segmentation mask.
\begin{equation}
    L= L_{cls} + L_{box}  + L_{mask} 
\end{equation}
where L\textsubscript{cls} is classification loss, L\textsubscript{box} is bounding box regression loss and L\textsubscript{mask} is mask loss. Dice coefficient performs better at class imbalanced problems. So we modified the L\textsubscript{mask} loss with Dice coefficient loss instead of binary cross-entropy loss. We observed that validation loss is converging smoothly with Dice coefficient loss. 

The segmentation from the CNN may contain some artifact which is not liver. To relieve this issue, some basic post-preparing was performed.

\section{Experiment and Results}  
Results of the automated liver segmentation are exhibited in Fig.4. Comparison of ground truth with the segmented liver is highly promising for obtaining high-performance metrics. The whole setup was implemented in Linux environment using NVIDIA GTX 1080 8 GB GPU on a system with 16GB  RAM and having Intel Core-i5 7\textsuperscript{th} generation @3.20GHz processor.  

\begin{figure}[ht]
    \centering
    \begin{center}
    \begin{tabular}{c} 
    \includegraphics[scale = 0.6]{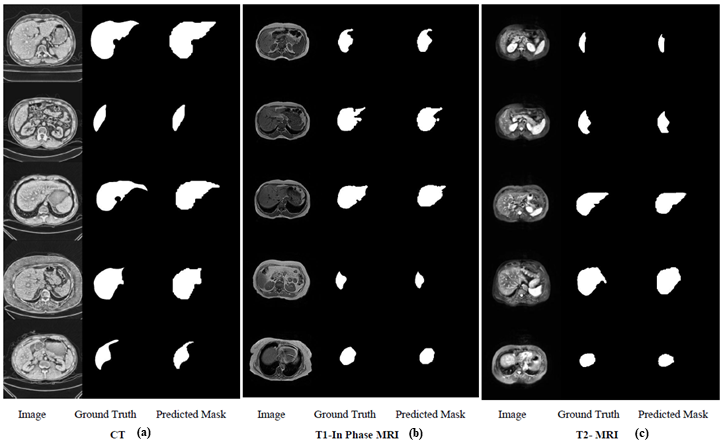}
    \vspace{-3mm}
    \end{tabular}
    \end{center}
    \caption{CT  and MR images with their respective ground truth and predicted output.}
    {\label{fig:example4}
    }
\end{figure} 

\subsection{Datasets}

The network training is run on a subset of the publicly available CHAOS challenge \footnote[3]{https://chaos.grand-challenge.org/} containing data sets from two different modalities.  The training is applied to 2183 axial CT slices(16 patients), 452 (4 patients) are used for validation. For MRI T2-weighted we applied training on 469 axial (7 patients) images and 80 (2 patients) images are used for validation. In the case of T1-weighted in-phase images, we use 316 axial (15 patients) images for training and 105 (5 patients) images for testing.

\subsection{Evaluation}
For each image, detections made by the model are compared to the ground truth label to evaluate the model performance given in the challenge for that image. We want to demonstrate the robustness, generalization, and scalability of our proposed method in this work. Table 1 provides the comparison between Mask R-CNN and HED-Mask R-CNN. Comparison with ground truth and segmented liver give rise to the assertion that our approach is highly promising for obtaining high-performance metrics. Dice coefficient of 0.9 above for both CT  and MRI images shows that our method works well.  

\begin{table}[ht]
\caption{Comparison of liver segmentation on CT and MRI scans} 
\vspace{-1mm}
\label{tab:Segmentation performance}
\begin{center}       
\begin{tabular}{ p{4cm} | p{1.5cm} | p{4cm} | p{1.5cm} }
\hline
\multicolumn{4}{c}{Quantitative comparison among Mask R-CNN and HED-Mask R-CNN} \\
\hline
CT Segmentation &  Dice  & MRI Segmentation  &  Dice  \\
\hline
Mask R-CNN   \newline (N=491)  & 0.90  &  Mask R-CNN  \newline (N=105 T1-weighted) & 0.80  \\
\hline
HED-Mask R-CNN    \newline (N=491)    & 0.94$\pm 0.03 $ & HED-Mask R-CNN   \newline (N=105 T1-weighted)    & 0.91$\pm 0.06$   \\
\hline
\multicolumn{4}{c}{
  \begin{minipage}{6.5cm}
    \tiny N = number of slices
  \end{minipage}
}\\
\end{tabular}
\end{center}
\end{table}

\section{Discussion}
\vspace{-1mm}
We propose a combination of HED (deep version) and Mask R-CNN network to improve the liver segmentation performance of CT/MRI imaging modality. We enhance the CT/MRI images which are shown in Fig. 2 to demonstrate the proficiency of the proposed joint deep network. It is seen from the Fig.2(b) and 2(d) that the edge information and contrast of the liver are enhanced than the original images. The enhanced images of CT/MRI are fed to the HED deep network to extract the edge map as shown in Fig. 3.

The segmented liver output of CT/MRI using the proposed joint network is shown in Fig. 4. The  Fig.4(a) depicts the segmented liver of CT images, Fig. 4(b) depicts the segmented liver of T1-in phase and Fig. 4(c) depicts the segmented liver of T2 MRI images. The segmentation results of our proposed joint deep network are compared with the ground truth extracted by the medical experts. It is observed that our proposed networks perform well in segmentation. The comparison of Dice value for Mask R-CNN and proposed HED-Mask R-CNN is tabulated in Table 1. It is observed from Table 1 the CT images without HED network obtained the Dice coefficient of 0.90 and with HED the Dice coefficient of 0.94. Similarly, for MRI images the Dice coefficient without and with HED network are 0.80 and 0.91 respectively. It is seen from Table 1 that HED network with the combination of Mask R-CNN increases the segmentation accuracy of the proposed network. 

The post-processing using graph cut method requires initial segmentation image with larger liver area and controlling parameters needs to be determined by multiple experiments in GIU-Net \cite{LiuZhe}. Whereas we present a framework, which is capable of a segmenting the liver with no post-processing strategy. Also generalized CNN \cite{Wang} relied on retrospective data to train and validate the multimodal CNN, in contrast our method takes the current data to train the model. Thus, the proposed joint deep networks outperform well for both CT and MRI modalities of images.

\section{Conclusions}
We demonstrate our strategy for liver segmentation in multimodal CT and  MRI images using HED-Mask R-CNN method in this work.The novelty of our work is in the use of edge map with Mask R-CNN with automatic features learning instead of just applying CNN for object segmentation. Importantly, the above framework obviates the need for liver segmentation significantly increasing robustness and accuracy as compared to stand-alone segmentation methods. Our method yields Dice 0.94 for CT and 0.91 for MRI images.  Our results on 491 CT slices and 105 slices demonstrate an impressive improvement over independent CNN based strategies and may give significant clinical estimations for liver segmentation.  We intend to apply our technique to segment liver from additional imaging modalities and all organs segmentation as well.

%
%
%

\end{document}